\documentclass[10pt,conference]{IEEEtran}
\usepackage{makecell}
\usepackage{times,amsmath,epsfig}
\usepackage{graphicx} 
\usepackage{subcaption} 
\usepackage{url}
\title{ML-based Real-Time Control at the Edge: An Approach Using hls4ml}
     \author{Omitted}%

\author{%
\parbox{\linewidth}{\centering{R. Shi{\small $~^{\#}$}, S. Ogrenci{\small $~^{\#}$}, J.M. Arnold, J.R. Berlioz, P. Hanlet, K.J. Hazelwood, \\M.A. Ibrahim, H. Liu{\small $~^{\#}$}, V.P. Nagaslaev, A. Narayanan{\small $~^{1}$}, D.J. Nicklaus, J. Mitrevski, G. Pradhan, \\A.L. Saewert, B.A. Schupbach, K. Seiya, M. Thieme{\small $~^{\#}$}, R.M. Thurman-Keup, N.V. Tran }}%
\vspace{1.6mm}\\
\fontsize{10}{10}\selectfont\itshape
$^{\#}$\,Northwestern University, Evanston, IL USA\\
\fontsize{9}{9}\selectfont\ttfamily\upshape
\vspace{-2.9mm}\\
\fontsize{10}{10}\selectfont\rmfamily\itshape
\,Fermi National Accelerator Laboratory, Batavia, IL USA\\
\fontsize{9}{9}\selectfont\ttfamily\upshape

\vspace{-2.9mm}\\
\fontsize{10}{10}\selectfont\rmfamily\itshape
$^{1}$\,also at Northern Illinois University, DeKalb, IL USA\\
\fontsize{9}{9}\selectfont\ttfamily\upshape

}
\begin{document}
\maketitle
\begin{abstract} 
This study focuses on implementing a real-time control system for a particle accelerator facility that performs high energy physics experiments. 
A critical operating parameter in this facility is beam loss, which is the fraction of particles deviating from the accelerated proton beam into a cascade of secondary particles. Accelerators employ a large number of sensors to monitor beam loss. The data from these sensors is monitored by human operators who predict the relative contribution of different sub-systems to the beam loss. Using this information, they engage control interventions. In this paper, we present a controller to track this phenomenon in real-time using edge-Machine Learning (ML) and support control with low latency and high accuracy. 
We implemented this system on an Intel Arria 10 SoC. Optimizations at the algorithm, high-level synthesis, and interface levels to improve latency and resource usage are presented. Our design implements a neural network, which can predict the main source of beam loss (between two possible causes) at speeds up to 575 frames per second (fps) (average latency of 1.74ms). The practical deployed system is required to operate at 320 fps, with a 3ms latency requirement, which has been met by our design successfully. 
\end{abstract}

%

\section{Introduction}
Real-time edge applications in experimental sciences are growing at an exponential rate, from high energy physics to astronomy, from material science to medical sciences. The main driver behind this trend is the advances in instrumentation across these domains. Scientific discovery using the data generated from the latest generation of instrumentation can only be supported by intelligent and adaptable computing systems. Real-time edge-ML computing is emerging as a promising solution to support experimental instrumentation with large scale sensing and data processing needs. 

High energy physics is one of the science domains at the forefront of developments in edge-ML hardware. A large and vibrant community of physicists, engineers, and computer scientists created the FastML initiative~\cite{noauthor_fastML}. A significant body of work on both ASIC and FPGA-based edge-ML is already formed, addressing many problems ranging from pre-processing detector data in particle accelerators~\cite{Govorkova2022-et, Kopciewicz:2800473, ieee_nss_talk_1_2020} to other high energy physics experiments~\cite{10.3389/fdata.2022.787421}. The domain of edge-ML hardware is growing in many other fields~\cite{10.3389/fdata.2022.787421, 10.1145/3572772}.   

Majority of the early work focused on deploying low latency edge-ML hardware for data processing.  Real-time control is a relatively new direction. Yet, there is a strong motivation to integrate low latency and small form factor systems into experiment control, where real-time response can improve both the reliability and overall quality of the ongoing experiment. In this paper, we address a control problem where an integrated FPGA SoC-based system is demonstrated to meet both the latency and quality requirements of a mission critical control task in a particle accelerator facility.

\begin{figure}[ht]
	\centerline{\psfig{figure=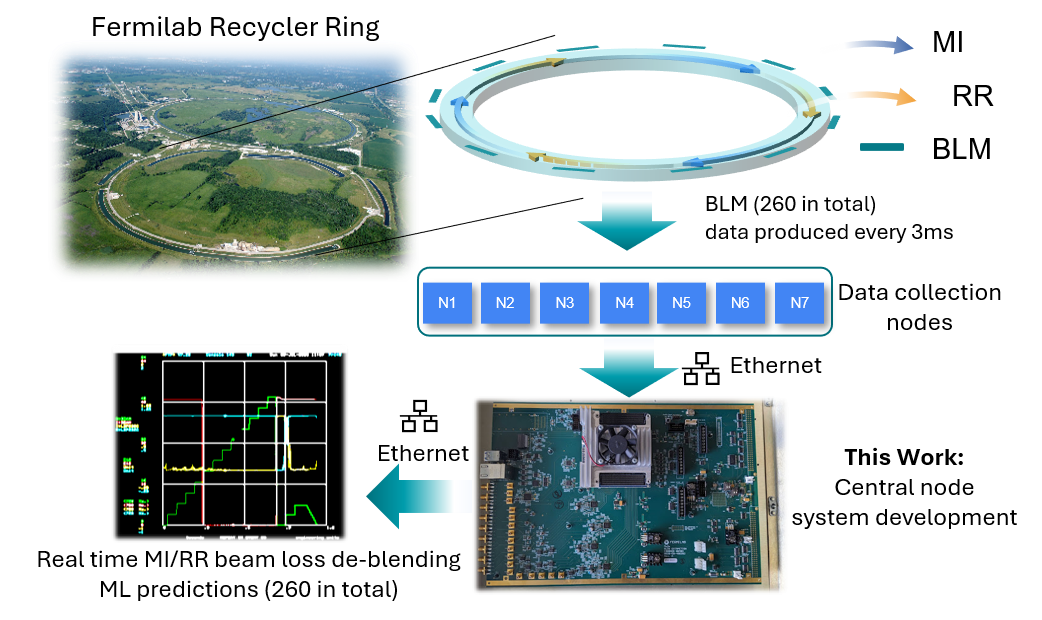,width=90mm} }
	\caption{Illustration of the accelerator tunnel that utilizes the beam loss de-blending system. There are 260 monitors distributed alongside the tunnel. They monitor the beam loss in real time. 
 }
	\label{fig:campus}
\end{figure}

Existing work on ML applications target different platforms such as CPUs, GPUs, ASICs, and FPGAs. Each has its own advantages. CPUs need the least development effort and they are suitable for low-cost applications with relaxed latency requirements. GPUs provide better balance between the development effort, scalability, and cost. However, they are only efficient when large batches of data is available for parallel processing. ASICs generally outperform other solutions in latency but are very expensive due to their long development cycle and fabrication costs with low flexibility for changing models post-deployment. FPGAs are suitable for low latency designs in $\mu$s to ms range and offer significant flexibility with their reconfigurable properties, as well as generally the best energy efficiency per inference.

\begin{table*}[ht]

\centering
    \caption{System Latency Comparison Across Multiple Models and Multiple Platforms for Sequential Inputs}     
    \label{tab:relatedwork}
    \begin{small}
    \begin{tabular}{|c|c|m{2.5cm}|c|c|c|c|c|c|m{1.2cm}|}
    \hline
    {\bfseries Work} & {\bfseries IP Core} & {\bfseries Typical Layers } & {\bfseries Params} & {\bfseries Precision} & {\bfseries ALMs} & {\bfseries Board} & {\bfseries Latency} & {\bfseries Data Tran.} & {\bfseries Tools}          \\
    \hline
    \makecell{VLSI'18~\cite{ma2018optimizing}}& CNN & Con2D, Pool & 7.59M & 16 bits& 161k & Arria 10  &  3.8 ms & DMA & RTL Compiler \\

    \hline
    \makecell{FPL'19~\cite{liu2019towards}}& U-Net & Con, Decon, Conct, Pool & ? & 8 bits& 250k & Arria 10  &  17.4 ms & DMA &  Verilog\\
    \hline
    \makecell{MLST'21~\cite{aarrestad2021fast}}& CNN & Dense, Con2D & 12,858 & 7 bits& 48k & PYNQ-Z2  &  0.17 ms & AXI DMA & hls4ml\\

    \hline
    \makecell{DATE'23~\cite{khandelwal2023quantised}}& MLP & Dense & ? & 4 bits& ? & ZCU104  &  0.12 ms & AXI & FINN \\
    \hline
    
    \makecell{This Work}& MLP & Dense & 100,102& 16 bits & 96k& Arria10  &0.31 ms & MM Bridge & hls4ml\\
    \hline
    \makecell{This Work}& U-Net & Dense, Con1D, UpSam, Pool, Conct & 134,434 & 16 bits& 224k & Arria10 & 1.74 ms & MM Bridge & hls4ml \\
    \hline
    \end{tabular}
    \end{small} 
    \vspace{-0.5cm}
\end{table*}

The control application we are targeting is required to meet ms-range latency requirements. Furthermore, a reconfigurable implementation is highly desirable as edge-ML models keep evolving and also in many experimental instrumentation the operating environment and data behavior can vary significantly over time, necessitating adaptation. Finally, FPGA SoCs offer a wide range of interfacing options. Particularly, the Hard Processor System (HPS) and its available utilities on the SoC enable practical and high performance Ethernet communication. Developers leverage the operating system deployed on the HPS to orchestrate the data communication with the experimental instrumentation system via Ethernet and the reconfigurable ML accelerator can be implemented as an IP on the FPGA. With the HPS performing some pre-processing and post-processing, the real-time data can be passed to the FPGA by the high-speed memory-mapped interface. In summary, FPGA SoCs stand out as a promising platform for implementing edge-ML hardware.

The real-time edge-ML problem we are tackling concerns performing an inference task with ms-latency to derive optimal operation parameters of a feedback controller. This controller is located within the particle accelerator tunnel of Fermilab, where high energy physics experiments are conducted. This particle accelerator tunnel (illustrated in Fig.~\ref{fig:campus}) is monitored in real-time to control an operation parameter called the beam loss, which is the fraction of the proton beam that does not contribute to the experiments. It represents waste of the critical high energy beam that, ideally, should be entirely used by experiments. Furthermore, a fraction of the beam protons deviating from their main path strike the accelerators resulting in ionizing radiation. This exposure can shorten the lifespan of the components, create a hazard for individuals working near that physical space, and in extreme cases, permanently damage the equipment. At the same time, being able to utilize the proton beam with minimal losses enables the scientists to maximize their experiment's efficiency and ensures that they can collect the maximum amount of data within the limited time they have access to the facility.    
In this particular setting, there are two particle accelerators operating jointly in the facility: the Main Injector (MI) and the Recycler Ring (RR). Controlling the beam loss involves collecting data from distributed monitors known as Beam Loss Monitors (BLMs)~\cite{berlioz2022synchronous}~\cite{ibrahim2023fpga}, transferring the sensor data to a centralized node, and processing the data by a pre-trained ML model to predict the primary source of beam loss between these two accelerators (MI or RR)~\cite{seiya2021accelerator, hazelwood2021real}.
The BLM hardware's digitizer poll rate is 3ms per decision, so that the lossy machine can be tripped off as soon as possible in order to control radioactivity~\cite{berlioz2022synchronous}. The complete process of mitigating the beam loss is also referred to as the beam loss de-blending. We implemented the beam loss de-blending system on the Achilles Arria 10 SoC FPGA board as the central node. Our specific contributions in this work are as follows:

\begin{itemize}
\item A full system is developed utilizing the reconfigurable and software-programmable components of the FPGA SoC and a resource optimal neural network architecture for the FPGA device is explored; 
\item We established an ML/High Level Synthesis (HLS) co-design methodology for resource optimization. Specifically, we used layer-based post-training quantization combined with reuse factor tuning to trade-off accuracy and resource utilization;
\item	As part of this co-design process, we performed a thorough latency analysis for the real-time application including all FPGA SoC components;
\item   We developed the first complete system with an HLS-based design flow integrating an open-source ML hardware design package (hls4ml) and the Intel HLS compiler.
\end{itemize}


\begin{figure*}[ht]
	\centerline{\includegraphics[width=\textwidth]{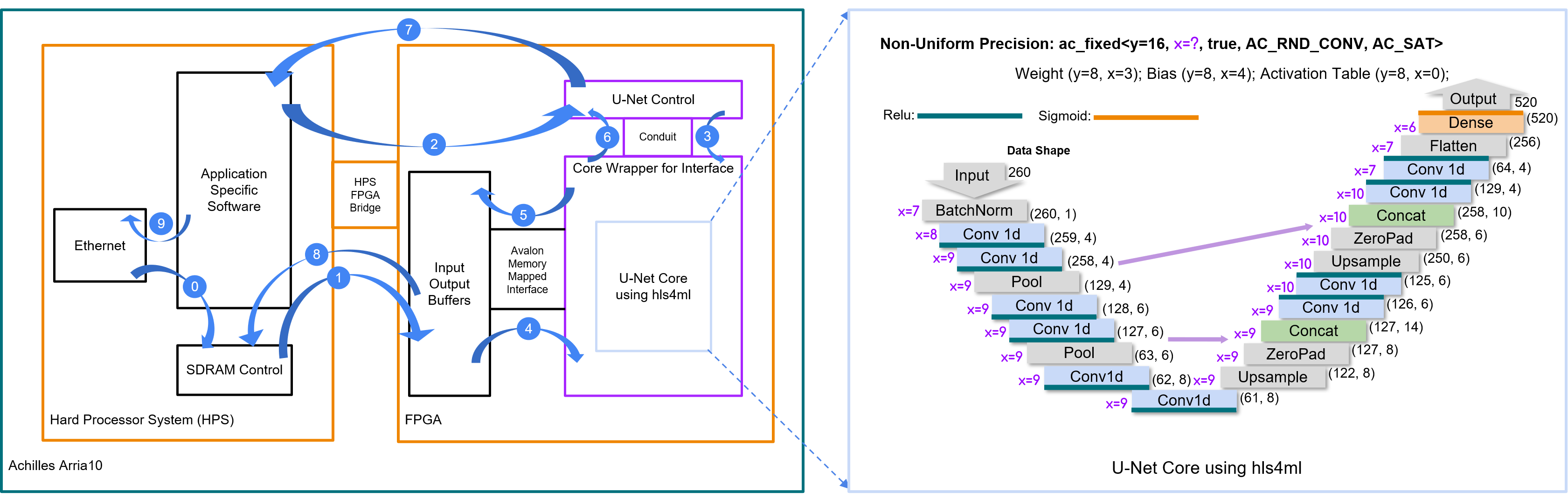}}
	\caption{The overall architecture on the central node and the deployed IP core.}
	\label{fig:socfpga}
\end{figure*}

\section{Related Work} \label{sec:rel}
To the best of the authors' knowledge, our solution is the first FPGA-based edge-ML control system addressing a real-time control task in an accelerator complex, meeting latency requirements with end-to-end equipment-controller interaction. A related real-time application in a particle accelerator setting that we are aware of~\cite{john2021real} uses a simple model (with approximately 34K parameters, while our model has over 134K parameters) and only meets a relaxed latency requirement (66 ms). Furthermore, it does not present the full system implementation and analysis. 

Table ~\ref{tab:relatedwork} presents other related works that implement neural networks on SoC FPGA boards for a variety of applications. As we indicated in the table, some solutions~\cite{ma2018optimizing, liu2019towards} deploy the neural network with a compiler that can translate to RTL or are directly described in RTL. In several cases they are using Direct Memory Access (DMA) for data transfer. DMA is tailored for transferring large chunks of data at a time and its use in these ML hardware solutions results in higher latencies. Quantization in HLS can help with latency and resources as shown in ~\cite{khandelwal2023quantised}. A prior work~\cite{aarrestad2021fast} also uses hls4ml in developing the IP core on an FPGA. This design has a similar latency to one of the two models we developed in our work (namely, similar to our MLP model implementation). However, our model has double the precision and contains 10-fold more trainable parameters compared to this prior work.




\section{Background} \label{sec:bg}
In this section, we review relevant concepts and assumptions concerning the ML model and the target hardware platform.
 
\subsection{ML Model} \label{subsec:mlmodel}
The central control node where the edge-ML hardware is deployed is designed for de-blending the beam loss. It receives inputs from seven BLM hubs distributed around the accelerator complex and sends outputs to the ACNET control system of the facility for further operations. The communication between these systems is performed via the Ethernet.

A pre-trained U-Net model~\cite{thieme2022semantic} serves as the kernel on the FPGA, as shown in Fig.~\ref{fig:socfpga}. It has an encoder-decoder structure with skip connections that can forward detailed features into the decoder. It can achieve accurate results with a relatively small training dataset. Our U-Net model contains layers shown in Fig.\ref{fig:socfpga}. The training of the U-Net model uses standardized floating point values. It contains 134,434 trainable parameters in total. In this figure each layer is annotated with its resource-aware custom layer-based precision (parameter x). We will elaborate further on this optimization in Section~\ref{sec:Opt}. Outputs of 260 beam loss monitors are provided as inputs to the model. Given that these 260 input values are received as an input array every 3 ms, the U-Net Intellectual Property (IP) is expected to produce 520 output values in an output array, each pair representing the probabilities of either one of the accelerators (MI or RR) being the primary source of the corresponding beam loss detected. Based on the output, the source with higher probability will be mitigated for that given time frame.  

We have also implemented a simpler MLP model to aid in our verification and early architecture exploration phase. It consists of two dense layers (128 and 518 nodes, respectively), and similar input size and output size. This model consists of 100,102 trainable parameters, 905 nodes, and uses a uniform precision of 16 bits.



\subsection{Target Platforms}



\begin{figure}[t]
	\centerline{\psfig{figure=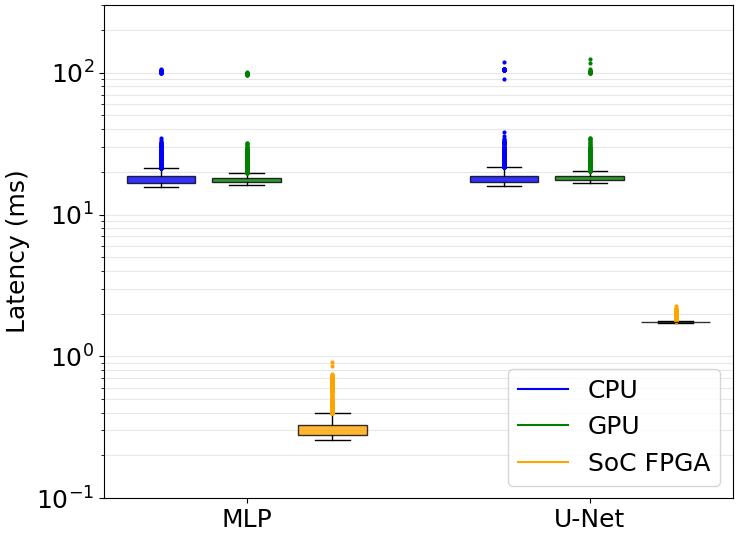,width=60mm} }
	\caption{System latency across multiple models and Platforms, batch size = 1.}
	\label{fig:latency}
        \vspace{-0.5cm}
\end{figure}
We first performed preliminary experiments to mimic the processing of 520 floating point values arriving every 3ms by creating the models described in Section~\ref{subsec:mlmodel}. We executed the Keras implementation of these models on a CPU and GPU system. As shown in Fig.~\ref{fig:latency}, the high CPU latency is mainly due to the long control and data path delays which cannot be customized for the needs of our specific model. The GPU performs well, with latencies in microseconds range, if we feed large batches of data to the GPU all at once. \textit{However, this is not the case for our control application. Sensor data arrives in small batches and in that case, we observe that the GPU performs similarly to the CPU.} This preliminary analysis further supported our choice to explore an FPGA SoC architecture, which we can customize with resource-aware optimization techniques. The following section introduces background for this FPGA SoC architecture.

\vspace{-0.2cm}

\section{Architecture} \label{sec:md}
\label{sec:page style}
In this section, we present the overall architecture of the control system and its mapping onto the Intel Arria 10 SoC. The two main components of the SoC are the reconfigurable fabric (we will refer to it as the FPGA) and the programmable processor (referred to as the Hard Processor System (HPS)).




\begin{figure*}[ht]
	\centerline{\includegraphics[width=\textwidth]{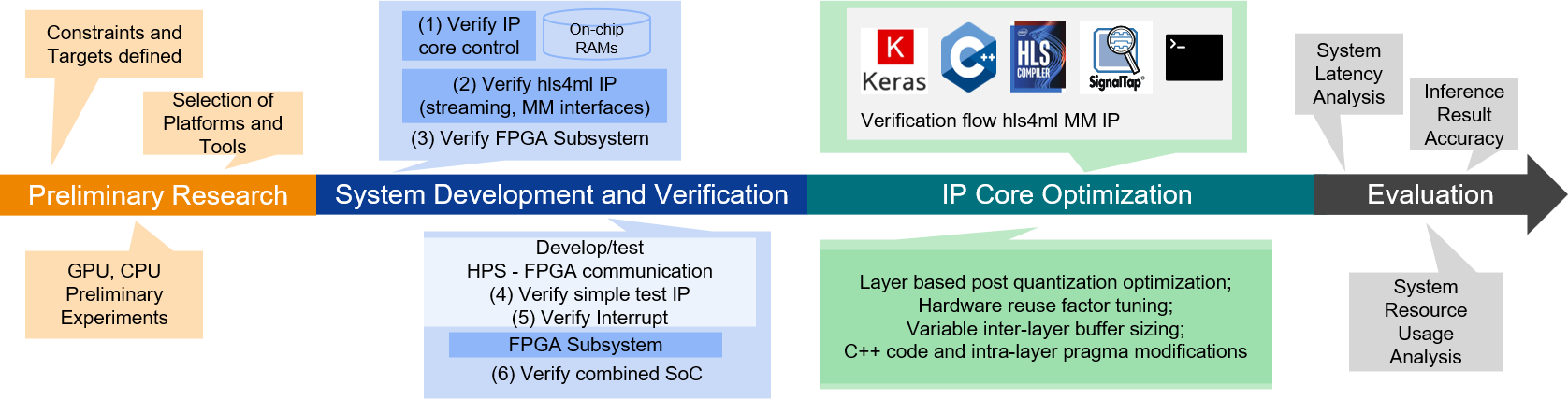}}
	\caption{The system development flow.}
	\label{fig:wflow}
\end{figure*}

\subsection{Architecture Overview} \label{sec:Archover}

The architecture of the central node is shown in Fig. \ref{fig:socfpga}. It deploys the U-Net IP core on the FPGA side and uses two on-chip RAMs as the input and output buffers. Following the steps labeled from \textit{0} to \textit{9} in the figure, once the input is ready in the SDRAM, the HPS will initiate a write to the input buffer through the FPGA-to-HPS bridges (\textit{Step 1}). When the write operation completes, the control will be notified by the bridge and the U-Net IP will be triggered (\textit{Step 2}). The U-Net IP signals the control when it finishes writing the results into the output buffer (\textit{Step 6}). The control then interrupts the HPS to let it read the result back to the SDRAM (\textit{Steps 7} and \textit{8}). This process from \textit{Step 1} to \textit{Step 8} takes 1.74 ms on average. (\textit{Step 9}) represents Ethernet communication off of the central node. 


\subsection{IP Core and System Development}

For the IP core development, we have created an HLS based design flow, using hls4ml~\cite{fahim2021hls4ml}, an open-source tool that converts the U-Net Keras model to a C++ project with HLS annotations. The development process is separated into several stages. Starting with the pre-trained U-Net model, we use hls4ml to produce the initial C++ program. With its default capabilities, hls4ml generates descriptions for IPs with streaming interfaces, hence, the IP can only consume data passively. We modified this default hls4ml interface by customizing the memory-mapped host interface for the IP core, enabling the IP to actively read/write from/to the external data buffer. The firmware of the U-Net IP core is then synthesized by the Intel HLS compiler using the C++ description generated by hls4ml. We also designed a dedicated control IP in HDL to handle the handshake between HPS and the U-Net IP.

We integrated all the components: the U-Net IP, the input/output buffers, the control IP, and performance counters to measure real latency through the platform designer facility of Quartus. This serves as a subsystem and is included in the Quartus project and compiled into a bitstream. We programmed the Achilles board with a prebuild Linux system on the HPS side using TFTP. Through the USB port on the board we are able to log into the system and run customized user space applications to interact with the U-Net IP and exchange data with the larger accelerator control facility system through the Ethernet.

\subsection{Verification Flow}

The verification of each stage is done by computing and comparing each stage's output with the expected Keras outputs. At the early stage of the system development, we tested sub-components separately and individually. Our tests were applied mainly to the following components: 1. the IP Core Control; 2. the hls4ml-generated streaming IP; 3. the systems on the FPGA side including on-chip RAM, the controller for the IP Core, and the IP core itself; 4. the Avalon memory mapped bridge testing with booted operating system accessing a simple component on the FPGA (we used a single adder) with the user space application; 5. the interrupt testing; and 6. testing of components (3), (4), and (5) combined with Signal Tap to diagnose potential problems.

\begin{itemize}
    \item Since (1) is relatively simple and written in HDL, we first verified its functionality on a smaller  Intel Cyclone V FPGA board. We created a test bench in VHDL with simple test control inputs. After completing the verification under our test cases through ModelSim simulations, this component has been set as ready for more tests in combination with the IP core. 
    \item The next step is to test the IP core. Out target IP core (the complete U-Net model) contained certain layer types which required co-design of the hls4ml tool together with the model to deliver the necessary design capabilities. During this phase, we started with a simpler model, a small MLP model, and verified each stage, including translation from the Keras code to the hls4ml output in C++, design of control logic in HDL for the streaming interface of the IP, and monitoring of the signals on Modelsim and SignalTap. 
    \item After verifying the hls4ml flow for the baseline MLP model and the associated default interface configurations, we then proceeded to extend hls4ml to add the capability of automatically creating a memory-mapped IP core interface which is required for seamless integration of the IP with the HPS component of the SoC. Utilizing the new interface we were able to combine the IP with the control logic and the on-chip RAM, to mimic the major part of the design residing on the FPGA side. We still tested this subsystem on a smaller Cyclone V board first, with a smaller IP. The only piece missing for this sub-system is a trigger. In the final deployment, the trigger is provided by the HPS side, to activate the IP and retrieve data. At this step, we simulated this trigger simply by changing a value in the on-chip RAM and using its positive edge as a trigger. Values in the on-chip RAM can be directly modified by the in-system memory content editor. The system is verified by monitoring real-time signals via the SignalTAP utility and validating the content in the on-chip RAM via the in-system memory content editor.
    \item Steps (4) and (5) are mainly concerned with verifying the physical board's functionalities. We used an example template project of the Achilles Arria 10 to verify tasks involved in steps (4) and (5). Finally, the complete system was debugged by observing the output from the user space application. \textit{Regarding the user space application development, care should be taken in the way the compiler is applying optimizations on the user space application code. Certain optimizations may result in the reordering of the program's instructions, which can incorrectly interfere with the start/end control and the timing of the inputs and outputs being written. This can result in incorrect operation.}
\end{itemize}

After the system is verified any future verification effort only needs to focus on the incremental updates of the IP alone and the rest of the surrounding interfaces and control logic is maintained, hence no verification effort needs to be spent on them. The outputs of different stages of the IP core development include the Keras program output, the C++ project quantization output generated by hls4ml, the co-simulation output from the Intel HLS compiler, simulation waveforms from ModelSim, the real-time FPGA signal waveforms measured on board using the SignalTap utility, and the customized user space application output after the system is deployed. We developed the corresponding testbench for each step and debugged through them. \textbf{We note that this verification flow is reusable for any other IP using the interfaces we have designed.}

\subsection{Optimization Methods} \label{sec:Opt}

We optimized the central node design at different stages to reduce resource usage while achieving good accuracy and satisfying the 3ms latency requirement. An important design aspect is the interface between the U-Net IP and the rest of the system. The interface of the IP needs to handle 260 16-bit inputs and 520 16-bit outputs. We chose Avalon Memory Mapped Host Interface to actively read/write from/to the external buffer while maintaining a relatively small port size. By reading in and writing out to the data buffer sequentially, we minimized Load-Store-Units and arbitration resources. Two dual-port RAMs serve as output buffers, a 16-bit data port is used for communication with the U-Net IP, and a 32-bit port is used for the communication with the HPS. We use a low overhead handshake protocol (AXI interface) between the HPS and the FPGA.

At the algorithm level, the model was trained with the original data with magnitudes ranging from 105,000 to 120,000 using a Batch Normalization Layer to perform the standardization. This resulted in poor accuracy given the tightly constrained range of the 16-bit resource-aware quantization. We then explored standardizing the data before training, which improved accuracy to the desired levels while using the same stringent quantization limits. We then pre-tested the architecture with randomized inputs and parameters and verified that the IP, its controller, and interfaces were functional on the FPGA SoC board. For the randomized U-Net model, all the parameters are between 0 and 1, while for the U-Net model trained with actual data using floating point values, the range of values for the parameters are much wider. That forced us to increase the precision to 18 bits to maintain similar accuracy. However, the resulting design exceeded the available number of ALUTs on the FPGA. In order to meet the resource constraints we re-evaluated \textbf{the maximum absolute output value generated inside each individual layer of the model}. Using this maximum, we calculated the required number of integer bits for each layer and adjusted each layer's precision individually while achieving a design with effective average precision equivalent to the homogeneous design with 16 bits precision. Fig.~\ref{fig:socfpga} has been annotated with the layer-based precision of each layer (x values listed in the figure). Fig.Table~\ref{tab:precision} presents the accuracy of the two outputs from our ML model and the corresponding resource utilization for different optimization choices we have explored, where \textit{layer-based precision achieved the best trade-off}. \textbf{The accuracy of the model is measured as a percentage of the cases where the quantized model output is close enough to the pre-trained model output. Each such comparison is classified as ``close enough'' when the difference between the two outputs is within 0.20 given the full output range is between 0 and 1.}

\begin{table}[ht]

\centering
    \caption{Optimization: Effect of Precision Customization on the U-Net Model}     
    \label{tab:precision}
    \begin{small}
    \begin{tabular}{|c|c|c|c|}
    \hline
    {\bfseries Strategy} & {\bfseries Accuracy}         & {\bfseries Accuracy}     & {\bfseries Resource}           \\
    & {\bfseries MI }         & {\bfseries RR}     & {\bfseries ALUTs}\\
    \hline
    \makecell{Uniform Precision\\ ac\_fixed$<$18, 10$>$}& 98.8\% & 99.3\% & 115\% \\
    \hline
    \makecell{Uniform Precision\\ ac\_fixed$<$16, 7$>$}& 16.7\% & 36.5\% & 22\% \\
    \hline
    \makecell{Layer-based Precision\\  ac\_fixed$<$16, x$>$}& 99.1\% & 99.9\% & 31\% \\
    \hline
    \end{tabular}
    \end{small} 
\end{table}

By utilizing hls4ml, we explored the HLS level optimization, which focuses on the reuse factor as the primary resource-latency trade-off. In the case of a neural network implementation, the reuse factor indicates the number of times a multiplier in a hidden layer is reused for different operations. The higher the reuse factor, the less parallel the implementation is, minimizing resource use while sacrificing latency. We also empirically optimized other architecture parameters such as the data buffer size to pursue resource trade-offs and perform deadlock mitigation. 

\begin{figure*}[t] 
    \centering
    \begin{subfigure}{.32\textwidth}
        \centering
        \includegraphics[width=\linewidth]{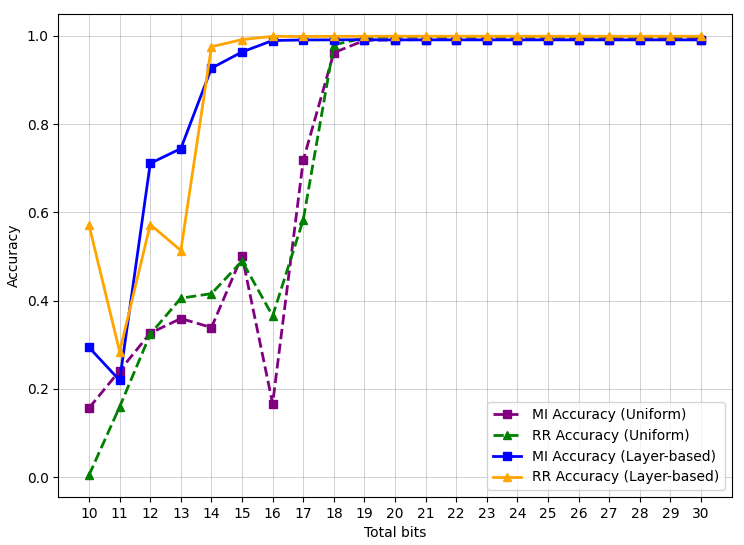}
        \caption{Change of Accuracy on MI and RR predictions as the number of total bits increases.}
        \label{fig:accuracy}
    \end{subfigure}%
    \hfill 
    \begin{subfigure}{.32\textwidth}
        \centering
        \includegraphics[width=\linewidth]{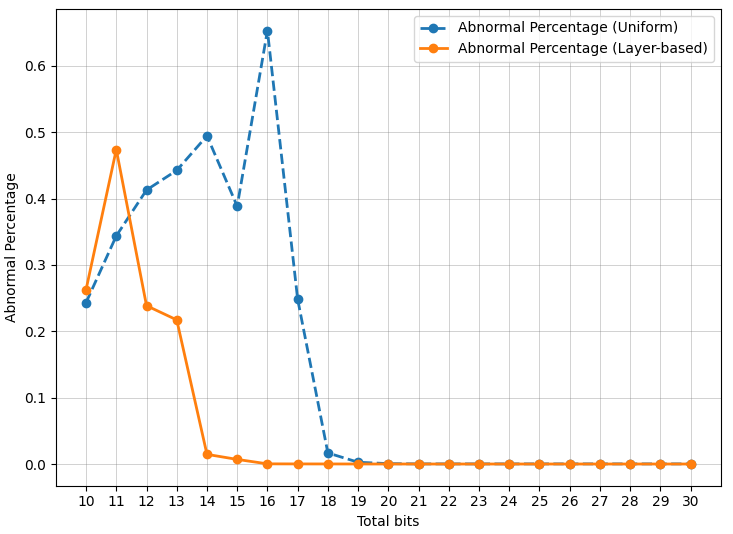}
        \caption{The number of outliers decreases as the number of total bits increase.}
        \label{fig:abnormal}
    \end{subfigure}%
    \hfill 
    \begin{subfigure}{.32\textwidth}
        \centering
        \includegraphics[width=\linewidth]{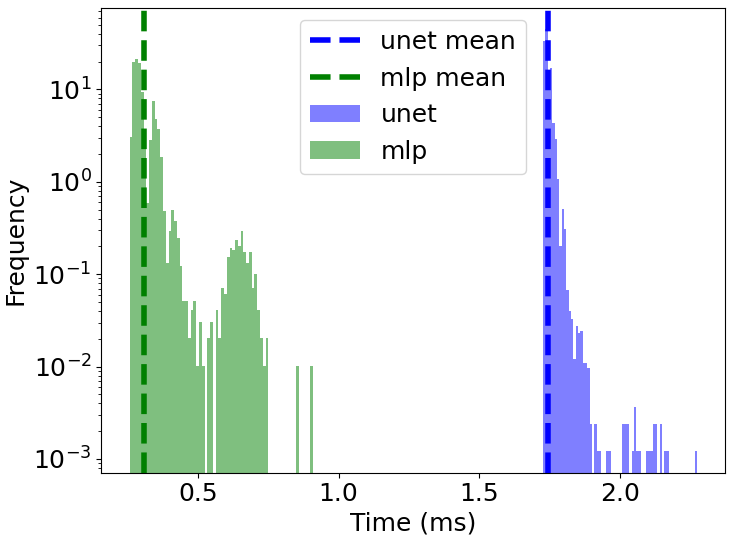}
        \caption{The distribution of system latency SoC FPGA (Steps 1-8 in Fig.3).}
        \label{fig:latency2}
    \end{subfigure}
    \vspace{1cm}
    \caption{Analysis of accuracy and timing.}
\end{figure*}

\textit{Finally, we note that our architecture template is highly flexible and can be adapted to other edge-ML computing problems. A majority of the central node system is transferable across different FPGA-SoC boards. The U-Net IP can be easily replaced by other IP cores as well, leveraging the general purpose interface wrapper we developed for hls4ml. For other real-time applications, the communication channels can also be easily reconfigured and further optimized.}

\setlength{\belowcaptionskip}{-10pt}
\begin{table}[h]
\begin{center}


    \caption{Model Summary}     
    \label{tab:summ}

    \begin{small}
    \begin{tabular}{|c|c|}
    \hline
    {\bfseries System Properties}     & {\bfseries U-Net Model}           \\
    \hline
    Trainable Parameters        &	134434		\\
    \hline
    Default Precision        & ac\_fixed$<$16, 7$>$ 	\\
    \hline
    Precision Strategy       & Layer-based 	\\
    \hline
    Default Reuse Factor      & 32		\\
    \hline
    Dense/Sigmoid Reuse Factor & 260		\\
    \hline
    Average System Latency      & 1.74ms		\\
    \hline
    FPGA U-Net Latency      & 1.57ms		\\
    \hline
    Logic Utilization      & 223674 (89\%)		\\
    \hline
    Total Registers      & 406123		\\
    \hline
    Total Pins      & 221 (37 \%)		\\
    \hline
    Total Block Memory Bits      & 25275808(58\%)		\\
    \hline
    Total RAM Blocks      & 1818 (85\%)		\\
    \hline
    Total DSP Blocks      & 273 (16\%)		\\
    \hline
    Total PLLs      & 3 (5 \%)		\\
    \hline
    \end{tabular}
    \end{small}
    
\end{center}
\end{table}

\section{Evaluation and Analysis} \label{sec:result}

Table \ref{tab:summ} presents the summary of the system implementation. Our main objective has been to achieve an ML module with high accuracy while meeting the resource and latency constraints. An important observation we made is that even for the same ML model architecture, the implementation of trained and untrained models can be very different. Across different training configurations, the range of data values being generated within the inner layers varies significantly. As we manage resource usage while trading off latency, we  need to increase the reuse factor of dense layers. The customization of the precision for each layer based on their maximum absolute values observed during profiling has been instrumental in achieving the optimal resource-latency trade-off.

Fig.5(a) presents the variation in accuracy between the original trained Keras model and quantized HLS code. The average difference is 0.025 for MI and 0.005 for RR, across 1,000 datasets (each dataset corresponds to one 260-input array). MI experiences more accuracy loss compared with RR. The reason behind this is as follows. The average magnitudes of the ML model's output (ranging from 0 to 1) is 0.17 for MI and 0.42 for RR. This means that generally the magnitude of outputs associated with RR is larger. Since quantization is performed based on the maximum absolute value, this causes the method to favor larger values and sacrifice the accuracy for smaller values. Therefore, RR benefits more from the quantization, thus, it experiences less loss of accuracy. There are still some infrequent outliers referred to as the abnormal points in Fig.5(b), which may occur because of inner layer overflows. We observed that half of these outliers could be mitigated by adding one extra bit to the integer part during quantization.





Fig.5(c) shows the overall latency distribution. It is measured starting from read access to the floating point input that resides in the SDRAM through the final step of the output being placed back into SDRAM. The average latency for U-Net and MLP is 1.74ms and 0.31ms, respectively. The measurements for the U-Net model range between 1.73ms and 2.27ms as shown in Fig.\ref{fig:latency}, for the MLP model this range is between 0.26ms and 0.91ms.  We observe that the overall latency of the U-Net model very rarely exceeds 2ms and 99.97\% of the cases the latency is below 1.9ms. The fluctuations above 2ms may originate from the task scheduling in the operating system.





\section{Conclusions}
We implemented an edge-ML computing solution to real-time control for high energy physics experiments on the Achilles Arria 10 board. The system includes a pre-trained U-Net model on the FPGA side and receives inputs in the form of floating point values communicated through the Ethernet. The system achieves a throughput of 575 frames per second, considering 260 beam loss sensor values at a given timestamp as one frame. The average latency of the complete system is 1.74ms (1.57ms contributed by the FPGA ML IP core) with a clock frequency of 100 MHz. This system has been fully implemented and tested on board and it is scheduled for final deployment in the particle accelerator facility of Fermi National Laboratories. 

\section{ACKNOWLEDGMENTS}

This work was performed at Northwestern with support from the Departments of Computer Science and Electrical and Computer Engineering. Operated by Fermi Research Alliance, LLC under Contract No.De-AC02-07CH11359 with the United States Department of Energy. Additional funding provided by DOE Grant Award No. LAB 20-2261~\cite{doe_lab_foa_2261}.


\bibliographystyle{IEEEtran}

\bibliography{IEEEabrv,IEEEexample}

\begin{thebibliography}{10}
\providecommand{\url}[1]{#1}
\csname url@samestyle\endcsname
\providecommand{\newblock}{\relax}
\providecommand{\bibinfo}[2]{#2}
\providecommand{\BIBentrySTDinterwordspacing}{\spaceskip=0pt\relax}
\providecommand{\BIBentryALTinterwordstretchfactor}{4}
\providecommand{\BIBentryALTinterwordspacing}{\spaceskip=\fontdimen2\font plus
\BIBentryALTinterwordstretchfactor\fontdimen3\font minus
  \fontdimen4\font\relax}
\providecommand{\BIBforeignlanguage}[2]{{%
\expandafter\ifx\csname l@#1\endcsname\relax
\typeout{** WARNING: IEEEtran.bst: No hyphenation pattern has been}%
\typeout{** loaded for the language `#1'. Using the pattern for}%
\typeout{** the default language instead.}%
\else
\language=\csname l@#1\endcsname
\fi
#2}}
\providecommand{\BIBdecl}{\relax}
\BIBdecl

\bibitem{noauthor_fastML}
``Fast machine learning lab,'' \url{https://fastmachinelearning.org/},
  accessed: 2023-07-25.

\bibitem{Govorkova2022-et}
E.~Govorkova \emph{et~al.}, ``Autoencoders on field-programmable gate arrays
  for real-time, unsupervised new physics detection at 40 {MHz} at the large
  hadron collider,'' \emph{Nature Machine Intelligence}, vol.~4, no.~2, pp.
  154--161, 2 2022.

\bibitem{Kopciewicz:2800473}
P.~Kopciewicz \emph{et~al.}, ``{Simulation and Optimization Studies of the LHCb
  Beetle Readout ASIC and Machine Learning Approach for Pulse Shape
  Reconstruction},'' CERN, Geneva, Tech. Rep.~18, 2021.

\bibitem{ieee_nss_talk_1_2020}
C.~Herwig \emph{et~al.}, ``{Design of a reconfigurable autoencoder algorithm
  for detector front-end ASICs},'' in \emph{IEEE Nuclear Science Symposium \&
  Medical Imaging Conference}, 2020.

\bibitem{10.3389/fdata.2022.787421}
\BIBentryALTinterwordspacing
A.~M. Deiana \emph{et~al.}, ``Applications and techniques for fast machine
  learning in science,'' \emph{Frontiers in Big Data}, vol.~5, 2022. [Online].
  Available:
  \url{https://www.frontiersin.org/articles/10.3389/fdata.2022.787421}
\BIBentrySTDinterwordspacing

\bibitem{10.1145/3572772}
\BIBentryALTinterwordspacing
H.~J. Damsgaard \emph{et~al.}, ``Approximation opportunities in edge computing
  hardware: A systematic literature review,'' \emph{ACM Comput. Surv.},
  vol.~55, no.~12, mar 2023. [Online]. Available:
  \url{https://doi.org/10.1145/3572772}
\BIBentrySTDinterwordspacing

\bibitem{ma2018optimizing}
Y.~Ma \emph{et~al.}, ``Optimizing the convolution operation to accelerate deep
  neural networks on fpga,'' \emph{IEEE Transactions on Very Large Scale
  Integration (VLSI) Systems}, vol.~26, no.~7, pp. 1354--1367, 2018.

\bibitem{liu2019towards}
S.~Liu \emph{et~al.}, ``Towards an efficient accelerator for dnn-based remote
  sensing image segmentation on fpgas,'' in \emph{2019 29th International
  Conference on Field Programmable Logic and Applications (FPL)}.\hskip 1em
  plus 0.5em minus 0.4em\relax IEEE, 2019, pp. 187--193.

\bibitem{aarrestad2021fast}
T.~Aarrestad \emph{et~al.}, ``Fast convolutional neural networks on fpgas with
  hls4ml,'' \emph{Machine Learning: Science and Technology}, vol.~2, no.~4, p.
  045015, 2021.

\bibitem{khandelwal2023quantised}
S.~Khandelwal \emph{et~al.}, ``Quantised neural network accelerators for
  low-power ids in automotive networks,'' in \emph{2023 Design, Automation \&
  Test in Europe Conference \& Exhibition (DATE)}.\hskip 1em plus 0.5em minus
  0.4em\relax IEEE, 2023, pp. 1--2.

\bibitem{berlioz2022synchronous}
J.~Berlioz \emph{et~al.}, ``Synchronous high-frequency distributed readout for
  edge processing at the fermilab main injector and recycler,'' \emph{arXiv
  preprint arXiv:2208.14873}, 2022.

\bibitem{ibrahim2023fpga}
M.~Ibrahim \emph{et~al.}, ``Fpga architectures for distributed ml systems for
  real-time beam loss de-blending,'' Fermi National Accelerator Laboratory
  (FNAL), Batavia, IL (United States), Tech. Rep., 2023.

\bibitem{seiya2021accelerator}
K.~Seiya \emph{et~al.}, ``Accelerator real-time edge ai for distributed systems
  (reads) proposal,'' \emph{arXiv preprint arXiv:2103.03928}, 2021.

\bibitem{hazelwood2021real}
K.~Hazelwood \emph{et~al.}, ``Real-time edge ai for distributed systems
  (reads): Progress on beam loss de-blending for the fermilab main injector and
  recycler,'' Fermi National Accelerator Lab.(FNAL), Batavia, IL (United
  States), Tech. Rep., 2021.

\bibitem{john2021real}
J.~S. John \emph{et~al.}, ``Real-time artificial intelligence for accelerator
  control: A study at the fermilab booster,'' \emph{Physical Review
  Accelerators and Beams}, vol.~24, no.~10, p. 104601, 2021.

\bibitem{thieme2022semantic}
M.~Thieme \emph{et~al.}, ``Semantic regression for disentangling beam losses in
  the fermilab main injector and recycler,'' Fermi National Accelerator
  Lab.(FNAL), Batavia, IL (United States), Tech. Rep., 2022.

\bibitem{fahim2021hls4ml}
F.~Fahim \emph{et~al.}, ``hls4ml: An open-source codesign workflow to empower
  scientific low-power machine learning devices,'' \emph{arXiv preprint
  arXiv:2103.05579}, 2021.

\bibitem{doe_lab_foa_2261}
\BIBentryALTinterwordspacing
{Department of Energy, Office of Science}. (2020) {Data, Artificial
  Intelligence, and Machine Learning at DOE Scientific User Facilities}.
  [Online]. Available:
  \url{https://science.osti.gov/-/media/grants/pdf/lab-announcements/2020/LAB_20-2261.pdf}
\BIBentrySTDinterwordspacing

\end{thebibliography}

\end{document}